\documentstyle[12pt]{article}
\def\lh{\ell_H}
\def\opt{{_{\rm opt}}}
\def\p{{\partial}}
\def\tr{\mathop{\rm tr}}
\def\H{{\cal H}}
\def\Re{\mathop{\rm Re}}
\title{Dynamical critical behavior in the integer quantum Hall effect}
\author{
Y. Avishai\\
\small Department of Physics, Ben Gurion University,\\
\small Beer-Sheva 84105, Israel\thanks{yshai@bgumail.bgu.ac.il}\\
\\
J.M. Luck\\
\small C.E.A., Service de Physique Th\'eorique,\\
\small 91191 Gif-sur-Yvette cedex, France\thanks{luck@spht.saclay.cea.fr}}
\date{}
\begin{document}
\maketitle
\begin{abstract}
We investigate dynamical scaling properties in the integer quantum Hall effect
for non-interacting electrons at zero temperature,
by means of the frequency-induced peak broadening
of the dissipative longitudinal conductivity $\sigma_{xx}(\omega)$.
This quantity is calculated numerically in the lowest Landau level
for various values of the Fermi energy $E$, of the frequency $\omega$,
and of the system size $L$.
Data for the width $W(\omega,L)$ of the peak
are analyzed by means of the dynamical finite-size scaling law
$W(\omega,L)\approx L^{-1/\nu}f\bigl(\omega L^z\bigr)$,
where $\nu$ is the static critical exponent of the localization length,
and $z$ is the dynamical exponent.
A fit of the data, assuming $\nu=2.33$ is known, yields $z=1.19\pm 0.13$.
This result indicates that the dynamical exponent in the
integer quantum Hall effect may be different from the
pertinent space dimension ($d=2$),
even in the absence of interactions between electrons.
\end{abstract}
\newpage
Dynamical scaling is well understood for usual critical phenomena,
in the vicinity of phase transitions driven by
thermal fluctuations~\cite{PC}.
Its validity for quantum critical phenomena at zero temperature,
such as the Anderson localization or the transitions between Hall plateaus,
is less firmly established.
The understanding of dynamical scaling has been one of the motivations
for investigating frequency-dependent response functions
of a two-dimensional electron gas in a magnetic
field~\cite{Powell,Kuchar,Goldberg}.
A peculiar physics shows up,
both in the regime of weak magnetic fields~\cite{Efetov,Imry}
and in the integer quantum Hall regime~\cite{Shahar}.
Henceforth we will concentrate on the latter situation.
A suitable quantity to consider in this context
is the real (dissipative) part of the frequency-dependent
(AC) longitudinal conductivity $\sigma_{xx}(\omega)$,
which has been measured~\cite{Shahar} and investigated theoretically~\cite{PS}.
One of the most striking features
of the quantum Hall effect~\cite{JM:P,JM:B,JM:H,JM:A}
is the vanishing of the static (DC) dissipative conductivity,
except right at the critical values $E_c$ of the Fermi energy $E$
(usually the centers of the Landau bands), where extended states appear
(hereafter we will use the difference $E-E_c$ in Fermi energy
as a measure of the distance to criticality,
instead of the magnetic field or the filling factor).
If we now take into account the effects of either temperature $T$,
or sample size $L$, or frequency $\omega$,
the dependence of the dissipative conductivity
$\sigma_{xx}(\omega,T,L,E)$ on the Fermi energy
is broadened from a delta function $\delta(E-E_c)$
to a peak of width $W(\omega,T,L)$.

In the following we shall investigate the scaling properties
of the frequency-induced width of the peak,
focusing, for simplicity, on the zero-temperature limit
and on the lowest Landau level.
In fact, the temperature-induced and frequency-induced peak broadening
have been argued to be governed by the same exponent~\cite{Sarma,Sondhi},
a comprehensive discussion of the combined effects of temperature and frequency
being given in footnote 30 of Ref.~\cite{Sarma}.

In order to relate the width $W(\omega,L)$
to the dynamical exponent $z$,
we recall that the localization length of the electronic states
diverges as $\xi(E)\sim\vert E-E_c\vert^{-\nu}$
as the Fermi energy $E$ approaches its critical value $E_c$.
The physics of this static localization transition has been extensively
studied~\cite{CC,Bodo,JM:Mi,HA,JM:HB,Zhang,Sarma}.
Numerical calculations consistently yield the critical exponent
$\nu\approx 2.33$ for a wide class of disordered potentials,
in remarkable agreement with a heuristic argument,
based on a semi-classical picture of percolating trajectories,
predicting $\nu=7/3$~\cite{JM:MIL,JM:A}.
The natural assumption of dynamical scaling is that the slowest processes
have a relaxation time of order $\tau\sim\xi^z\sim\vert E-E_c\vert^{-z\nu}$,
with $z$ being the dynamical exponent of the system~\cite{PC}.
For a large but finite sample,
the AC longitudinal conductivity at zero temperature is expected to
depend on frequency only through the product $\omega\tau$,
and to obey finite-size scaling~\cite{JM:FSS}, namely to depend
on the sample size only through the ratio $L/\xi$.
We are thus led to the following scaling law for the conductivity
\begin{equation}
\sigma_{xx}(\omega,L,E)
\approx{e^2\over h}F\Bigl(\vert E-E_c\vert\,L^{1/\nu},\omega L^z\Bigr).
\label{sca}
\end{equation}
The critical DC conductivity at $E=E_c$ assumes the universal value
$\sigma_c=(e^2/h)F(0,0)$.
Accurate numerical calculations~\cite{Bhatt,JM:GB} yield $F(0,0)\approx0.5$,
confirming thus earlier estimates~\cite{Hikami,CC}.
The corrections to the scaling law~(\ref{sca}) have been found~\cite{JM:GB}
to be proportional to $L^{-\eta}$, with $\eta\approx 1.63$.

The following scaling law for the width,
\begin{equation}
W(\omega,L)\approx L^{-1/\nu}f\bigl(\omega L^z\bigr),
\label{FSS}
\end{equation}
is a consequence of Eq.~(\ref{sca}).
For an infinite electron gas, the scaling laws (\ref{sca}) and (\ref{FSS})
respectively read
\begin{equation}
\sigma_{xx}(\omega,E)\approx{e^2\over h}
G\Bigl(\vert E-E_c\vert\,\omega^{-\kappa}\Bigr)
\end{equation}
and
\begin{equation}
W(\omega)\sim\omega^\kappa,
\label{kappasca}
\end{equation}
with
\begin{equation}
\kappa={1\over z\nu}.
\end{equation}
In the above expressions, $F$, $f$, and $G$ are universal scaling functions
in the usual sense.

The experiment~\cite{Shahar} yields $\kappa\approx 0.41$
for spin-split Landau levels, which for $\nu\approx 2.33$
implies a dynamical exponent $z\approx 1$.
The result $z=1$ has been given a theoretical explanation~\cite{PS}
in terms of variable-range hopping of interacting electrons.
In the case of non-interacting electrons,
where the physics is a priori simpler,
a commonly accepted argument~\cite{JM:CCL,JM:H},
recalled in the discussion below, yields $z=2$.

The aim of the present work is a direct numerical evaluation
of the dynamical exponent $z$ in the lowest Landau level,
within the framework of the integer quantum Hall effect
for non-interacting electrons at zero temperature.
The dissipative conductivity $\sigma_{xx}(\omega,L,E)$
and the width of its peak $W(\omega,L)$
are calculated for systems of various sizes $L$.
The data are then analyzed by means of the dynamical finite-size scaling
law~(\ref{FSS}).
The static and dynamical exponents $\nu$ and $z$
can thus be determined simultaneously, at least in principle.

We have considered a two-dimensional electron gas
confined in a square of side $L$, with periodic boundary conditions.
In this geometry it is convenient to define the
vector potential in the Landau gauge,
namely ${\bf A}=(A_x,0)$ with $A_x=-Hy$,
where $H$ is the magnetic field strength.
The magnetic length $\lh=(\hbar c/eH)^{1/2}$
is assumed to be much smaller than the sample size $L$.
The impurities are modeled as a lattice of $N^2$ Gaussian potentials
of spatial width $\Delta$, centered at the points
$(x_n,y_m)=\big(-a/2+(n-1)a,-a/2+(m-1)a\big)$,
with $n,m=-N/2+1,\cdots,N/2$, assuming $N=L/a$ is an even integer.
The random potential is written as
\begin{equation}
V(x,y)=\sum_{nm}{v_{nm}\over2\pi\Delta^2}
\exp\left(-\frac{(x-x_n)^2+(y-y_m)^2}{2\Delta^2}\right).
\label{potential}
\end{equation}
The impurity strengths $v_{nm}$ are independent random numbers,
which have been chosen for convenience to be uniformly distributed
between $-w/2$ and $w/2$, so that $\langle v_{nm}\rangle=0$
and $\langle v_{nm}v_{n'm'}\rangle=(w^2/12)\delta_{nn'}\delta_{mm'}$.
Since the distribution of the random potential is even,
observable quantities such as the density of states or the conductivity tensor
are symmetric with respect to the center of the Landau band $(E_c=0)$.

The interaction \ref{potential} is to be 
diagonalized within the subspace of functions
belonging to the lowest Landau level, that are periodic
in both co-ordinates $x$ and $y$, with period $L$~\cite{AA}.
The number of these mutually orthogonal basis functions is equal to the
extensive degeneracy of the Landau level, namely $K=L^2/(2\pi\lh^2)$.
Here we use the set of functions $\Phi_k(x,y)$ 
suggested in Eq. (1) of ref. \cite{YHL},
where $k$ is an integer between $-K/2+1$ and $K/2$.
The matrix elements $V_{kk'}$ of the potential \ref{potential}
can be calculated explicitly. The pertinent electron Hamiltonian 
is thus represented in $k$-space by a $K\times K$ matrix.
It has eigenvalues $E_\alpha$, with $\alpha=1,\cdots,K$,
and eigenvectors $\vert\alpha\rangle$,
with $\langle k\vert\alpha\rangle=c_\alpha(k)$.
The eigenfunctions in configuration space then read
$\langle x,y\vert\alpha\rangle=\Psi_\alpha(x,y)=\sum_kc_\alpha(k)\Phi_k(x,y)$.

Let us now make a choice of parameters.
We require that there be one impurity per magnetic area $\lh^2$,
and relate $a$ and $\Delta$ in such a way that the reduced overlap integral
between the potentials of two adjacent impurities
be neither too small, nor too close to unity.
The value $e^{-1/2}$ for the overlap yields $a=\lh=\Delta\sqrt2$.
These parameters correspond to $c_i=\pi$ and $d=\lh$,
in the notation of previous works~\cite{AA}.
In the following, the system size $L$ is given in units of $a$.
The strength $w$ of disorder is the only energy scale in the problem,
since the Hamiltonian $\H=V$ is projected onto the lowest Landau level.
For convenience, $w$ is fixed by requiring
that the second moment of the density of states be equal to unity.
An analytical evaluation of $\tr\H^2$,
too lengthy to be reported here, fixes $w=a^2\sqrt{48\pi}$.
With this choice of parameters, the density of states is not far from
a normalized Gaussian curve, and essentially vanishing for $\vert E\vert>3$.

Given the eigenstates $\Psi_\alpha(x,y)$ it is possible,
at least in principle, to evaluate dynamical quantities
by means of the linear-response formalism.
Strictly speaking, however, the matrix elements of the current
between states belonging to the same Landau level vanish.
Yet, as was shown in Ref.~\cite{AndoAC},
response functions can be evaluated from the knowledge of
states belonging to a given Landau level only,
by using in the Kubo formula the velocity of the guiding centers,
${\bf v}=(v_x,v_y)=\lh^2/\hbar\Bigl(\p V/\p y,-\p V/\p x\Bigr)$.
Note also that the use of the Kubo formula
for the evaluation of the conductivity
of finite systems requires some care.
Following the procedure as underlined in Ref.~\cite{AndoAC},
within linear-response theory, the real part of the dissipative
conductivity at the Fermi energy $E$ is given by
\begin{equation}
\Re\sigma_{xx}(\omega)=\frac{\pi e^2}{\hbar\omega L^2}\sum_{E_\beta<E<E_\alpha}
\Bigl\vert\langle\alpha\vert v_x\vert\beta\rangle\Bigr\vert^2
\Bigl[\delta(\omega-\omega_{\alpha\beta})
+\delta(\omega-\omega_{\beta\alpha})\Bigr],
\label{cond}
\end{equation}
where $\hbar\omega_{\alpha\beta}=E_\alpha-E_\beta$.
For any finite system, the spectrum of the Hamiltonian is discrete,
and the use of Eq.~(\ref{cond}) will represent the conductivity as a finite
sum of delta functions.
Since response functions are expected to have a smooth dependence
on frequency in the thermodynamical limit,
a proper smoothing procedure has to be applied.
Even for a finite sample of size $L\gg\lh$,
we are mainly interested in frequency ranges smaller than the
Landau-level spacing, but much larger than the average level spacing
within a single Landau band
(this is the domain of frequencies for which Eq.~(\ref{cond}) applies).
Thus, following an earlier work~\cite{AndoAC}, we smooth the delta functions
entering Eq.~(\ref{cond}) over a frequency interval $\Omega$,
such that $\hbar\Omega$ is a few
times the average level spacing in the lowest Landau band.
The resulting conductivity is finite as $\omega\to 0$,
yielding thus the DC conductivity.
To be on the safe side, we have restricted our calculations to the range
$\hbar\omega\le 0.24$, which is about 8 percent of the effective bandwidth.
We have checked our numerical procedure by applying it to the same system as
used in Ref.~\cite{AndoAC}, using the same parameters,
and the results concerning the conductivity
as a function of frequency were in agreement, within error bars.

The occurrence of a frequency-induced broadening is demonstrated in Figure~1,
where the longitudinal conductivity
$\sigma_{xx}(\omega,E)$ is plotted as function of the Fermi energy $E$,
for a large system size $(L=80)$, and several values of the frequency.
The DC conductivity exhibits a sharp decrease as a function of $E$,
which is a crossover effect between the system size $L$
and the correlation length $\xi(E)$.
For large $E$, we have $L\gg\xi$, so that the conductivity is very small.
In the present case, with $L=80$,
the DC conductivity essentially vanishes near $E\approx 0.8$,
while the density of states takes appreciable values up to $E\approx 3$.
In the opposite regime, when the energy approaches the band center,
we have $L\ll\xi$, so that the DC conductivity increases above zero.
The limit as $E\to 0$ is independent of the system size
(within error bars), as expected from scaling,
and compatible with the universal conductivity
of $0.5e^2/h$, recalled below Eq.~(\ref{sca}).
The most salient qualitative features of the AC conductivity
are also apparent on Figure~1: as frequency is increased,
the width of the peak increases, while its height slightly decreases.

In order to pursue the analysis at a quantitative level,
a convenient definition of the frequency-induced width $W(\omega,L)$ is needed.
We have chosen the following one,
which is suitable for a numerical treatment:
\begin{equation}
W(\omega,L)^2=\frac{\sum_iE_i^2\sigma_{xx}(\omega,L,E_i)}
{\sum_i\sigma_{xx}(\omega,L,E_i)},
\label{widthdef}
\end{equation}
where the $E_i$ are the energies at which the conductivity is
evaluated numerically.

We have thus obtained data for $W(\omega,L)$
for system sizes ranging from $L=20$ to $L=80$,
and frequencies ranging from $\hbar\omega=0$ to $\hbar\omega=0.24$.
The data have been fitted to the scaling law~(\ref{FSS})
by searching for which values of the critical exponents
$\nu$ and $z$ an optimal collapse is observed on plotting
$y=WL^{1/\nu}$ against $x=\hbar\omega L^z$.
This search has been facilitated by a fitting of
the rescaled data to the following M\"obius-quadratic form:
$y=f(x)=a+b/(x-x_0)+c/(x-x_0)^2$.
For fixed values of $\nu$ and $z$, the parameters $a,b,c,x_0$ are determined,
and the $\chi^2$ per degree of freedom is evaluated.
The statistical errors on the data are larger than
the systematic discrepancy between the true scaling function $f(x)$
and the above fitted form over the limited range of available data.
It is worth noticing that the asymptotic behavior $f(x)\sim x^\kappa$
for $x\gg 1$, needed in order to match Eq.~(\ref{kappasca}),
is not reproduced by the above parametrization.
This illustrates the power of finite-size scaling:
data in the asymptotic scaling region are not needed,
but a good data collapse in the crossover region $(x\sim 1)$
is sufficient to yield the critical exponents.

Figure~2 shows the data collapse, with its analytical fit,
for the following choices of the critical exponents:
\begin{itemize}
\item (a) we fix $\nu$ to its commonly accepted value $\nu=2.33$,
and choose the optimal value $z=z\opt=1.19$, so that the $\chi^2$ is minimal,
namely $\chi^2=1.26$ per degree of freedom.
By requiring that the $\chi^2$ does not exceed twice this minimal value,
we obtain the estimate $z=1.19\pm 0.13$.
\item (b) if we now fix $\nu=2.33$ and $z=1$,
we obtain $\chi^2=4.62$ per degree of freedom,
and a fit of somewhat lower quality than (a).
\item (c) as a cross-check, $z=1$ is now fixed,
yielding $\nu=\nu\opt=2.58$, and $\chi^2=1.65$ per degree of freedom.
Similarly to case (a), we obtain the estimate $\nu=2.58\pm 0.23$,
marginally including $\nu=2.33$.
\item (d) for comparison, $z=2$ is now fixed,
yielding $\nu=\nu\opt=1.60$, and $\chi^2=1.70$ per degree of freedom,
as well as the estimate $\nu=1.60\pm 0.13$.
Although the $\chi^2$ is not bad, the fit looks poorer than (a) and (c),
and the value of $\nu$ is definitely too far from $\nu=2.33$.
\end{itemize}

The present analysis, yielding $z=1.19\pm 0.13$,
does not concur with the standard viewpoint that $z=2$
in systems of non-interacting electrons.
Although more work is needed in order to draw a more definitive conclusion,
we can give, for the time being, the following qualitative discussion.
The result $z=d$ for non-interacting electrons~\cite{JM:CCL,JM:H},
recalled above, is based on the assumption that the only pertinent energy scale
is the mean level spacing, namely
\begin{equation}
\Delta E\sim\frac{1}{L^d\rho(E_c)}.
\label{DOS}
\end{equation}
This argument may, however, be questioned at the critical point $E=E_c$.
Indeed, within a renormalization-group approach,
an effective long-distance action is derived, describing the critical states.
It may well turn out that the density of states
entering the estimate (\ref{DOS}) only describes these critical states,
and is thus a renormalized quantity,
having no simple relation to the bare density of states $\rho(E_c)$.
This picture also shows up in a natural way
in the Chalker-Coddington network model~\cite{CC},
admitting ``a single extended state'' at the critical energy.
If the renormalized density of states has an anomalous dimension $y>0$,
we have $\Delta E\sim L^{y-d}$, hence $z=d-y=2-y$ for $d=2$.
Another argument in favor of the plausibility of an anomalous dimension
is provided by the supersymmetric approach
to the quantum Hall effect~\cite{JM:E}.
Within this framework, a term of the form $\omega{\rm Str}(Q\Lambda)$
is added to the critical action, driving the system away from criticality.
We have again $z=2-y$, where $y$ is the anomalous
dimension of the above operator,
to be determined within a renormalization-group scheme.
There is no obvious reason why this procedure will not lead to a
non-trivial anomalous dimension.
Indeed the absence of an anomalous dimension $(y=0)$
would mean that the two-point correlation of the operator $Q$ is
essentially given by the metallic diffusion propagator $1/(i\omega-Dq^2)$,
which has no reason to hold in the critical region.
Let us mention that anomalous dynamical exponents have been found recently
in several field-theoretical models~\cite{JM:Ts},
related to the physics of two-dimensional non-interacting electron systems.

To summarize, the present investigation has demonstrated numerically
the validity of the dynamical finite-size scaling law~(\ref{FSS})
for the frequency-induced width of the peak
of the longitudinal conductivity in the lowest Landau level.
From a quantitative viewpoint we find $z=1.19\pm 0.13$,
and suggest arguments in favor of the possibility
of an anomalous dynamical exponent $z\ne d$,
even for a system of non-interacting electrons.

\subsection*{Acknowledgments}
Y.A. would like to thank the SPhT at Saclay for their warm hospitality.
Discussions with Y. Goldschmidt, E. Shimshoni,
and B. Shapiro are gratefully acknowledged.
We are most grateful to A.M. Tsvelik for invaluable comments and suggestions.
The research of Y.A. is partially
supported by a grant from the Israeli Ministry of Science and Arts,
and partially by the Israeli Academy of Science.

\newpage
\section*{Figure captions}

\noindent{\bf Figure 1:}
Plot of the longitudinal dissipative conductivity $\sigma_{xx}(\omega,E)$
as a function of Fermi energy $E$,
for several values of the frequency $\omega$.
The system size is $L=80$.

\medskip
\noindent{\bf Figure 2:}
Finite-size scaling analysis of the width $W(\omega,L)$
of the frequency-induced peak of the longitudinal conductivity.
The scaling variables are $x=\hbar\omega L^z$ and $y=WL^{1/\nu}$.
Data for various system sizes are plotted,
together with the analytical fit described in the text.
(a) $\nu=2.33$ fixed, $z$ fitted to $z\opt=1.19$;
(b) $\nu=2.33$ and $z=1$ fixed;
(c) $z=1$ fixed, $\nu$ fitted to $\nu\opt=2.58$;
(d) $z=2$ fixed, $\nu$ fitted to $\nu\opt=1.60$.

\newpage
\begin{thebibliography}{99}
\bibitem{PC}
B.I. Halperin and P.C. Hohenberg, Phys. Rev. {\bf 177}, 952 (1969);
Rev. Mod. Phys. {\bf 49}, 435 (1977).
\bibitem{Powell}
T.G. Powell, R. Newbury, A.P. Long, C. McFadden, H.W. Myron, and M. Pepper,
J. Phys. C {\bf 18}, L 497 (1985);
T.G. Powell, R. Newbury, A.P. Long, H.W. Myron, and M. Pepper, Surf. Sci.
{\bf 170}, 173 (1986).
\bibitem{Kuchar}
F. Kuchar, R. Meisels, G. Weimann, and W. Schlapp, Phys. Rev. B {\bf 33} (RC),
2965 (1986).
\bibitem{Goldberg}
B.B. Goldberg, T.P. Smith, M. Heiblum, and P.J. Stiles, Surf. Sci. {\bf 170},
180 (1986).
\bibitem{Efetov}
O. Viehweger and K.B. Efetov, J. Phys. Cond. Matt. {\bf 2}, 7049 (1990);
Phys. Rev. B {\bf 44}, 1168 (1991).
\bibitem{Imry}
Y. Imry, Phys. Rev. Lett. {\bf 71}, 1868 (1993).
\bibitem{Shahar}
L.W. Engel, D. Shahar, C. Kurdak, and D.C. Tsui, Phys. Rev. Lett. {\bf 71},
2638 (1993).
\bibitem{PS}
D.G. Polyakov and B.I. Shklovskii, Phys. Rev. Lett. {\bf 70}, 3796 (1993);
Phys. Rev. B {\bf 48}, 11167 (1993).
\bibitem{JM:P}
R.E. Prange and S.M. Girvin (eds.), {\it The Quantum Hall Effect}
(Springer Verlag, New York, 1987).
\bibitem{JM:B}
D. Belitz and T.R. Kirkpatrick, Rev. Mod. Phys. {\bf 66}, 261 (1994).
\bibitem{JM:H}
B. Huckestein, Rev. Mod. Phys. {\bf 67}, 357 (1995).
\bibitem{JM:A}
A. Hansen, E.H. Hauge, and F.A. Maa{\o}, {\it Localization and Delocalization
in the Integer Quantum Hall Effect}, Springer Lecture Notes in Physics,
J. Duarte (ed.), to appear.
\bibitem{Sarma}
D. Liu and S. Das Sarma, Phys. Rev. B {\bf 49}, 2677 (1994).
\bibitem{Sondhi}
S.L. Sondhi and S.A. Kivelson, {\it Finite-Temperature
Scaling of the Transition Between Quantum Hall Plateaus} (preprint, 1995).
\bibitem{CC}
J.T. Chalker and P.D. Coddington, J. Phys. C {\bf 21}, 2665 (1988).
\bibitem{Bodo}
B. Huckestein and B. Kramer, Phys. Rev. Lett. {\bf 64}, 1437 (1990);
B. Huckestein, Europhys. Lett. {\bf 20}, 451 (1992).
\bibitem{JM:Mi}
B. Mieck, Europhys. Lett. {\bf 13}, 453 (1990);
Z. Phys. B {\bf 90}, 427 (1993).
\bibitem{HA}
H. Aoki, in {\it High Magnetic Fields in Semiconductor Physics III}
(G. Landwehr (ed.), Springer Verlag, Berlin, Heidelberg, 1992).
\bibitem{JM:HB}
Y. Huo and R.N. Bhatt, Phys. Rev. Lett. {\bf 68}, 1375 (1992).
\bibitem{Zhang}
S.C. Zhang, S. Kivelson, and D.H. Lee, Phys. Rev. Lett. {\bf 69}, 1252 (1992).
\bibitem{JM:MIL}
G.V. Mil'nikov and I.M. Sokolov, J.E.T.P. Lett. {\bf 48}, 536 (1988).
\bibitem{JM:FSS}
J. Cardy (ed.), {\it Finite-Size Scaling} (North-Holland, Amsterdam, 1988);
V. Privman (ed.), {\it Finite-Size Scaling and Numerical Simulation
of Statistical Systems} (World Scientific, Singapore, 1990).
\bibitem{Bhatt}
Y. Huo, R.E. Hetzel, and R.N. Bhatt, Phys. Rev. Lett. {\bf 70}, 481 (1993).
\bibitem{JM:GB}
B.M. Gammel and W. Brenig, Phys. Rev. Lett. {\bf 73}, 3286 (1994).
\bibitem{Hikami}
S. Hikami, Phys. Rev. B {\bf 29}, 3726 (1984);
S. Hikami and E. Br\'ezin, J. Phys. (Paris) {\bf 46}, 2021 (1985);
S. Hikami, Prog. Theor. Phys. {\bf 76}, 1210 (1986).
\bibitem{JM:CCL}
J.T. Chalker, J. Phys. C {\bf 20}, L 493 (1987);
J. Phys. C {\bf 21}, L 119 (1988).
\bibitem{AA}
T. Ando, J. Phys. Soc. Jpn. {\bf 52}, 1740 (1983);
J. Phys. Soc. Jpn. {\bf 53}, 3126 (1984);
H. Aoki and T. Ando, Phys. Rev. Lett. {\bf 54}, 831 (1985);
T. Ando and H. Aoki, J. Phys. Soc. Jpn {\bf 54}, 2238 (1985);
H. Aoki, Phys. Rev. B {\bf 33}, 7310 (1986);
Surf. Sci. {\bf 196}, 107 (1988);
T. Ando, Phys. Rev. B {\bf 40}, 5325 (1989);
Phys. Rev. B {\bf 40}, 9965 (1989);
Y. Ono, T. Ohtsuki, and B. Kramer, J. Phys. Soc. Jpn. {\bf 58}, 1705 (1989).
\bibitem{YHL}
D. Yoshioka, B.I. Halperin, and P.A. Lee, Phys. Rev. Lett. {\bf 50},
1219 (1983).
\bibitem{AndoAC}
T. Ando, J. Phys. Soc. Jpn. {\bf 53}, 3101 (1984).
\bibitem{JM:E}
K.B. Efetov, Adv. Phys. {\bf 32}, 53 (1983).
\bibitem{JM:Ts}
A.W.W. Ludwig, M.P.A. Fisher, R. Shankar, and G. Grinstein, Phys. Rev. B {\bf
50}, 7526 (1994);
A.A. Nersesyan, A.M. Tsvelik, and F. Wenger, Phys. Rev. Lett. {\bf 72}, 2628
(1994);
Nucl. Phys. B {\bf 438}, 561 (1995);
C. Mudry, C. Chamon, and X.G. Wen, Nucl. Phys. B {\bf 466}, 383 (1996);
J.S. Caux, I.I. Kogan, and A.M. Tsvelik, Nucl. Phys. B {\bf 466}, 444 (1996).
\end{thebibliography}
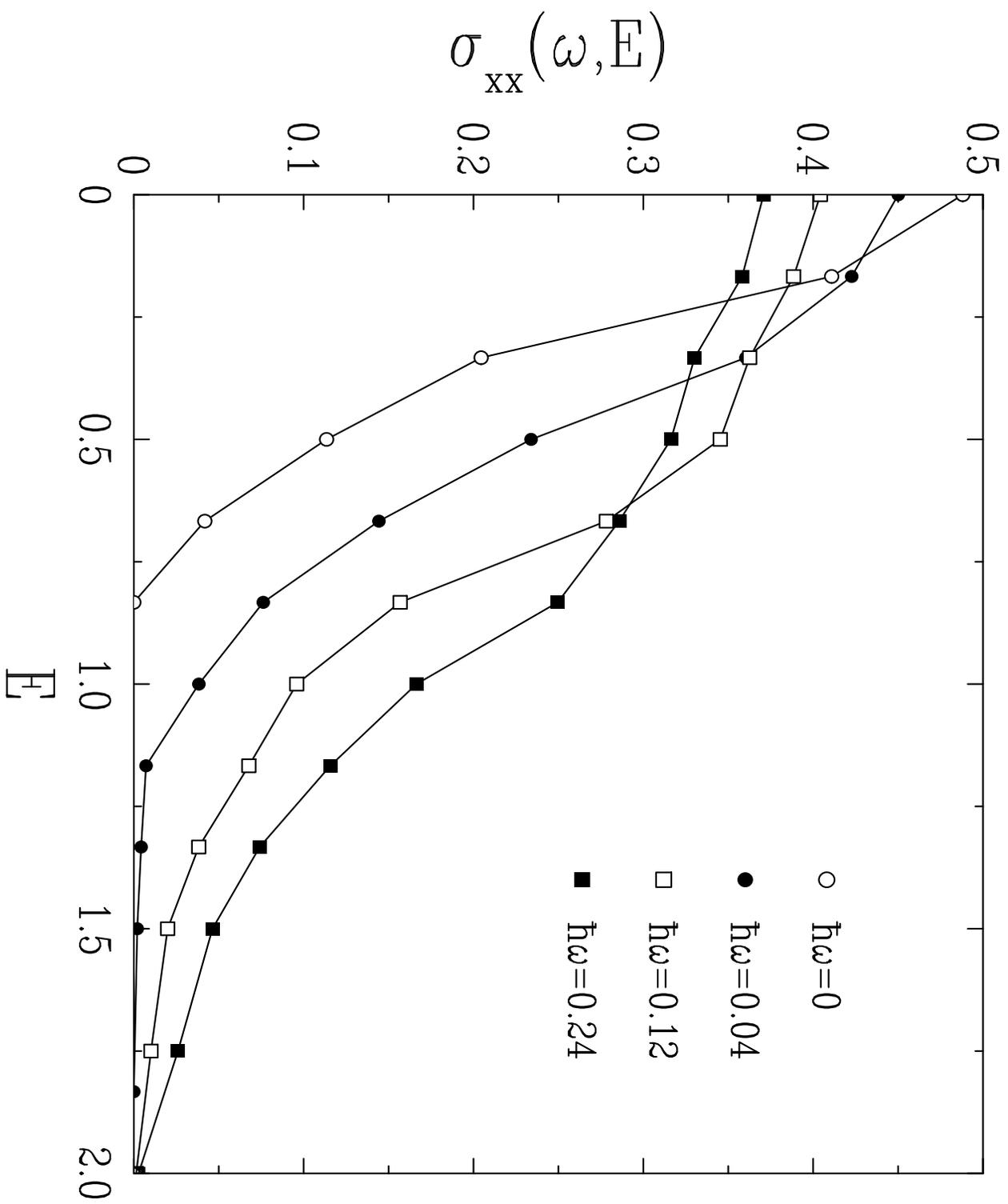
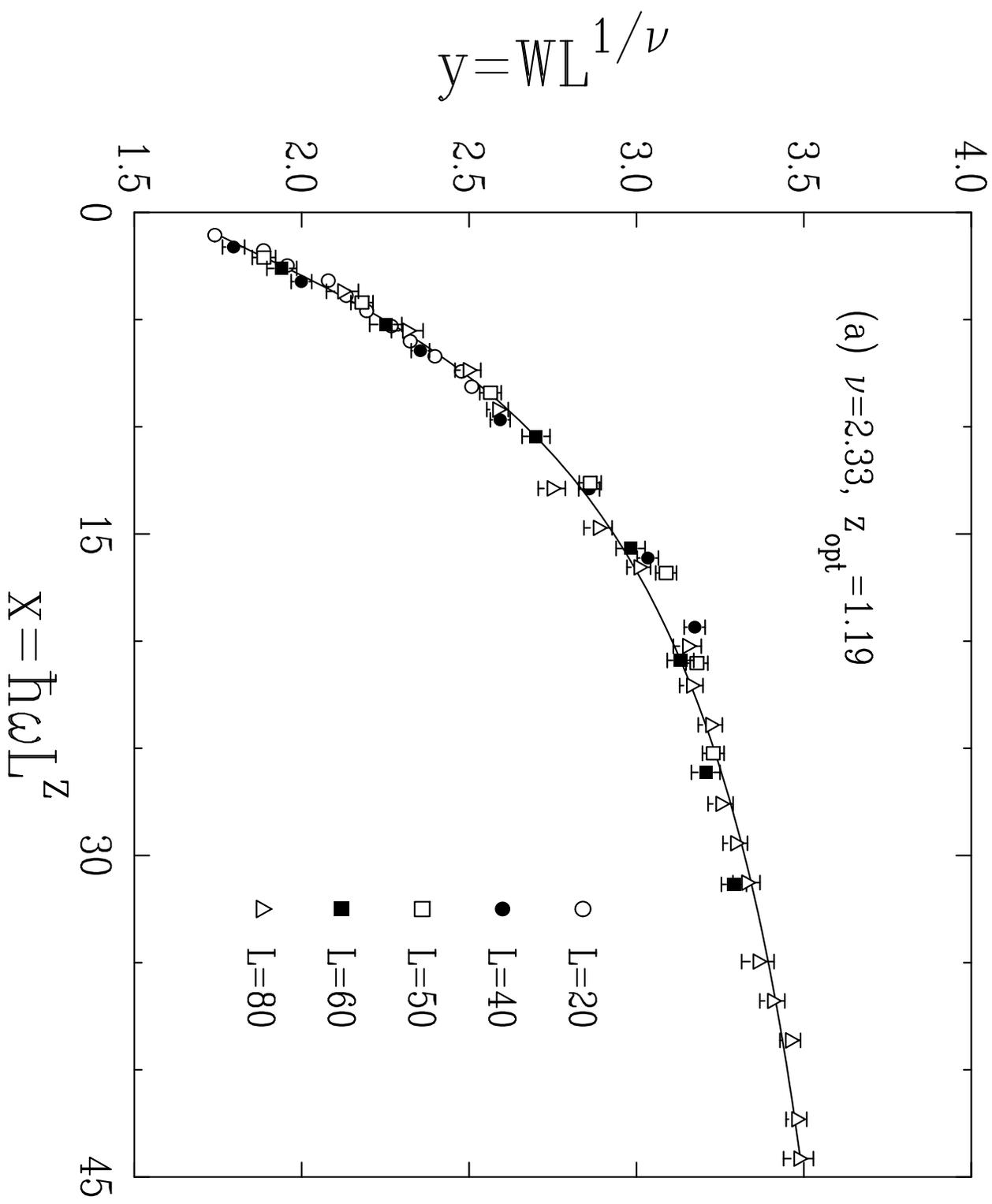
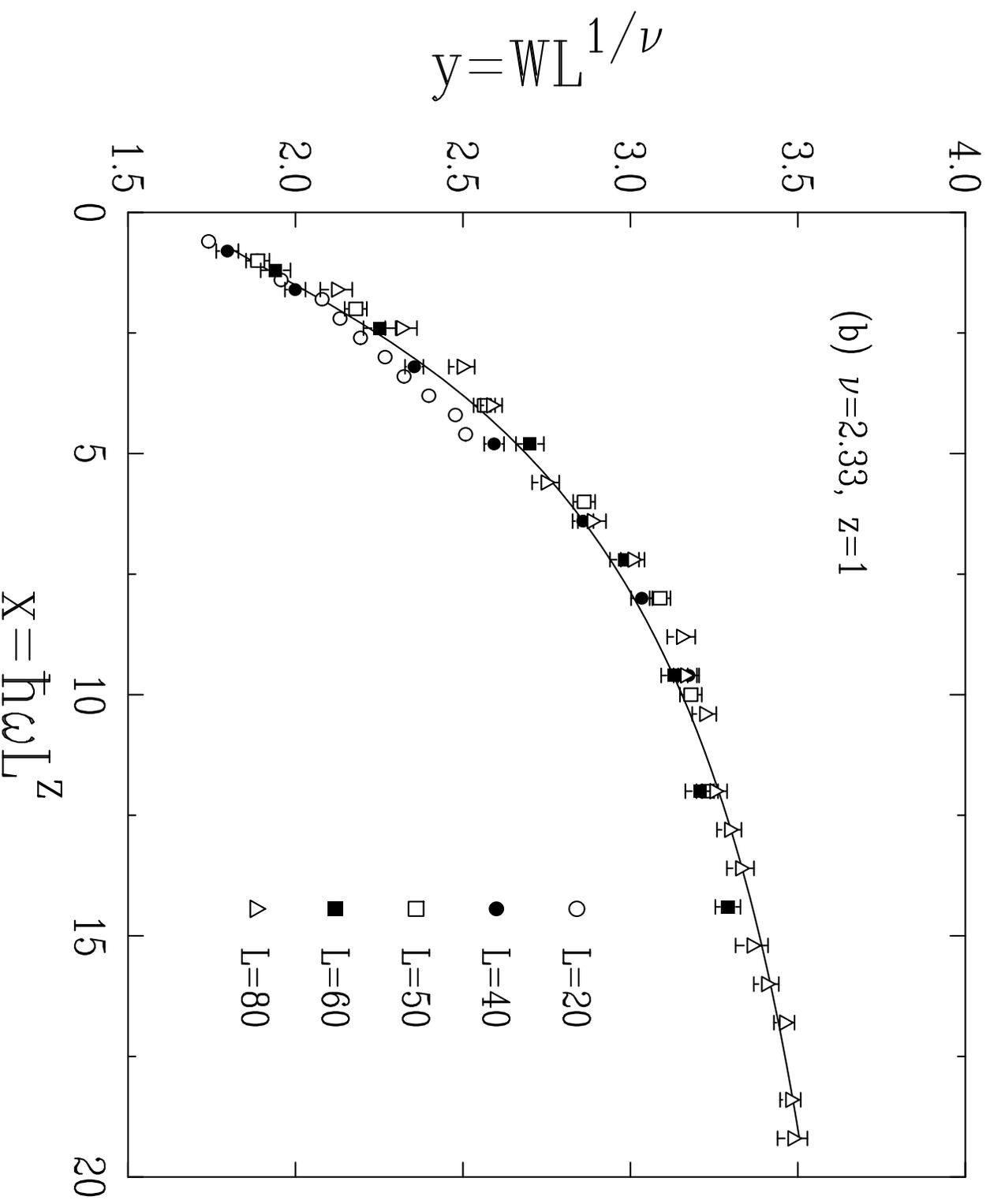
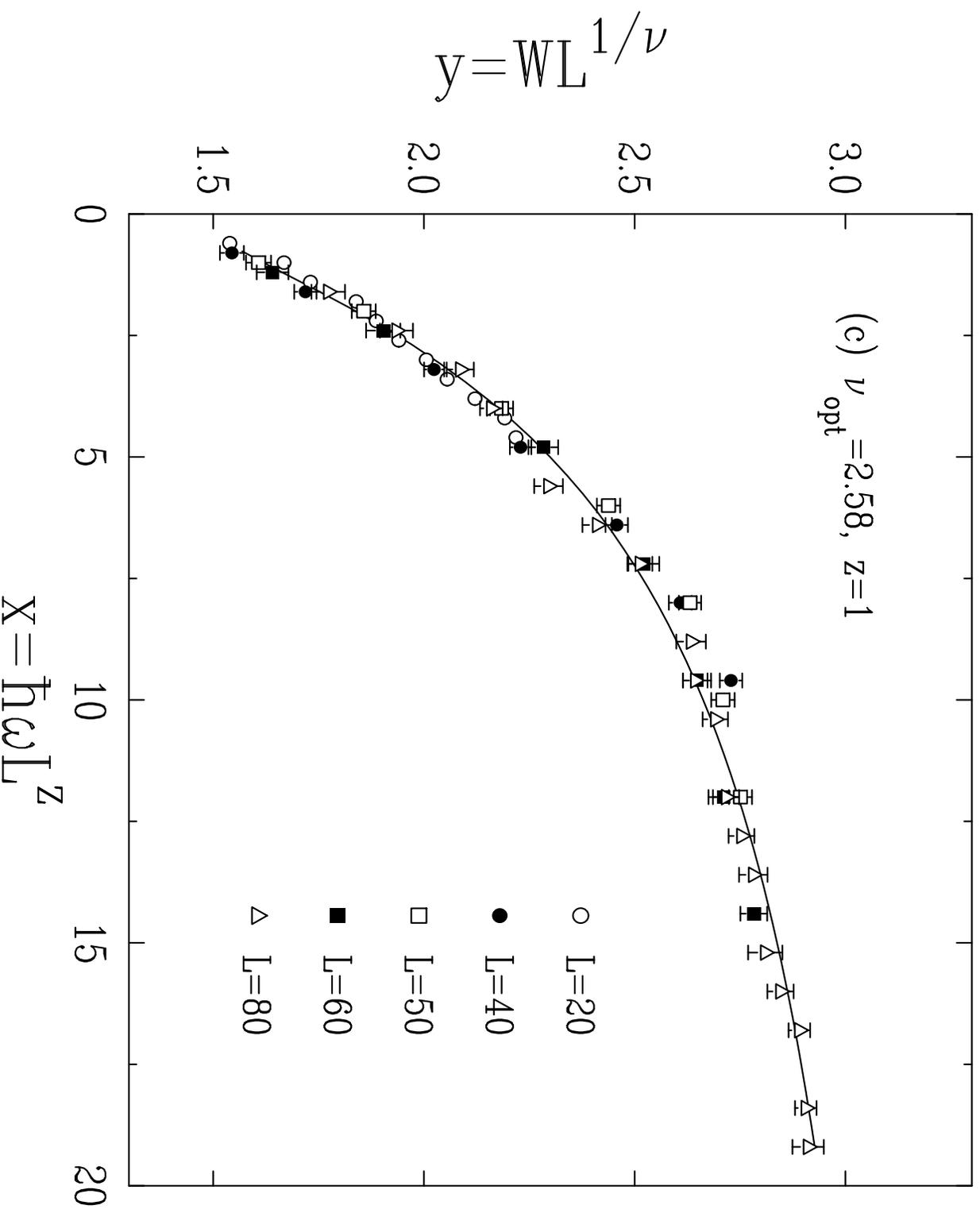
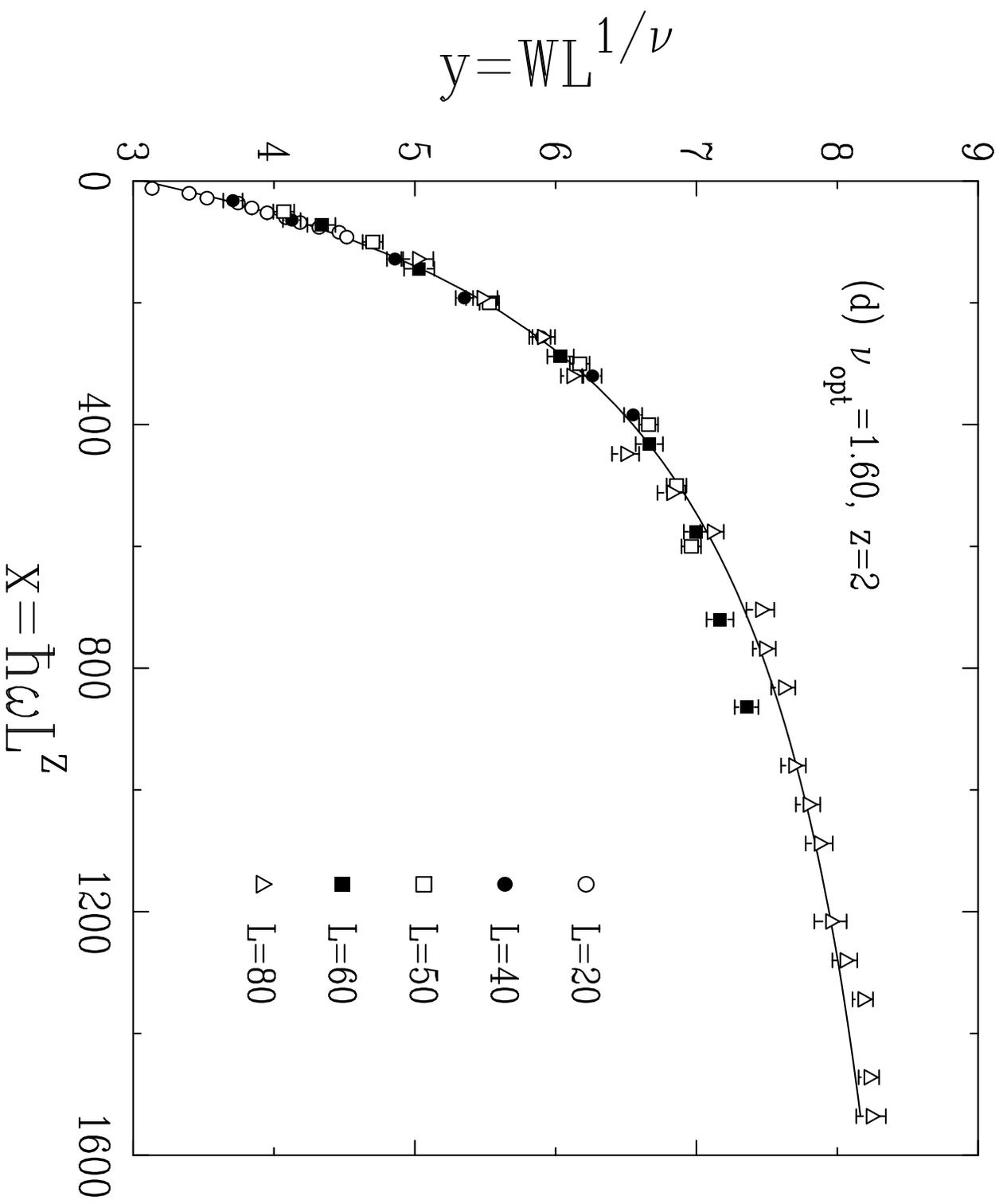
\end{document}